\documentclass[twocolumn]{emulateapj}
\usepackage{amsmath}

\newcommand{\etal}{et~al.}
\newcommand{\mum}{\ensuremath{\mu \mathrm{m}}}
\newcommand{\obj}{HERMES J105751.1+573027}
\newcommand{\shortobj}{HLSW-01}
\begin{document}
\title{DISCOVERY OF A MULTIPLY LENSED SUBMILLIMETER GALAXY IN EARLY
  HerMES {\it HERSCHEL}/SPIRE\footnote{{\it Herschel} is an ESA space
    observatory with science instruments provided by European-led
    principal Investigator consortia and with important participation
    from NASA.  In addition, some of the data presented herein were
    obtained at the W.M. Keck Observatory, which is operated as a
    scientific partnership among the California Institute of
    Technology, the University of California and the National
    Aeronautics and Space Administration. The Observatory was made
    possible by the generous financial support of the W.M. Keck
    Foundation.} DATA} 
\shorttitle{A MULTIPLY LENSED SUB-MM GALAXY}

\author{A.~Conley\altaffilmark{1},
A.~Cooray\altaffilmark{2,3},
J.D.~Vieira\altaffilmark{3},
E.A.~Gonz\'alez~Solares\altaffilmark{4},
S.~Kim\altaffilmark{2},
J.E.~Aguirre\altaffilmark{5},
A.~Amblard\altaffilmark{2},
R.~Auld\altaffilmark{6},
A.J.~Baker\altaffilmark{7},
A.~Beelen\altaffilmark{8},
A.~Blain\altaffilmark{3},
R.~Blundell\altaffilmark{9},
J.~Bock\altaffilmark{3,10},
C.M.~Bradford\altaffilmark{3,10},
C.~Bridge\altaffilmark{3},
D.~Brisbin\altaffilmark{11},
D.~Burgarella\altaffilmark{12},
J.M.~Carpenter\altaffilmark{3},
P.~Chanial\altaffilmark{13},
E.~Chapin\altaffilmark{14},
N.~Christopher\altaffilmark{15},
D.L.~Clements\altaffilmark{16},
P.~Cox\altaffilmark{17},
S.G.~Djorgovski\altaffilmark{3},
C.D.~Dowell\altaffilmark{3,10},
S.~Eales\altaffilmark{6},
L.~Earle\altaffilmark{1},
T.P.~Ellsworth-Bowers\altaffilmark{18},
D.~Farrah\altaffilmark{19},
A.~Franceschini\altaffilmark{20},
D.~Frayer\altaffilmark{21},
H.~Fu\altaffilmark{3},
R.~Gavazzi\altaffilmark{22},
J.~Glenn\altaffilmark{1,18},
M.~Griffin\altaffilmark{6},
M.A.~Gurwell\altaffilmark{9},
M.~Halpern\altaffilmark{14},
E.~Ibar,\altaffilmark{23},
R.J.~Ivison\altaffilmark{24,25},
M.~Jarvis\altaffilmark{25},
J.~Kamenetzky\altaffilmark{18},
M.~Krips\altaffilmark{17},
L.~Levenson\altaffilmark{3,10},
R.~Lupu\altaffilmark{5},
A.~Mahabal\altaffilmark{3},
P.D.~Maloney\altaffilmark{1},
C.~Maraston\altaffilmark{26},
L.~Marchetti\altaffilmark{20},
G.~Marsden\altaffilmark{14},
H.~Matsuhara\altaffilmark{27},
A.M.J.~Mortier\altaffilmark{16},
E.~Murphy\altaffilmark{3,28},
B.J.~Naylor\altaffilmark{10},
R.~Neri\altaffilmark{17},
H.T.~Nguyen\altaffilmark{10,3},
S.J.~Oliver\altaffilmark{19},
A.~Omont\altaffilmark{21},
M.J.~Page\altaffilmark{29},
A.~Papageorgiou\altaffilmark{6},
C.P.~Pearson\altaffilmark{30,31},
I.~P{\'e}rez-Fournon\altaffilmark{32,33},
M.~Pohlen\altaffilmark{6},
N.~Rangwala\altaffilmark{1},
J.I.~Rawlings\altaffilmark{29},
G.~Raymond\altaffilmark{6},
D.~Riechers\altaffilmark{3,34},
G.~Rodighiero\altaffilmark{20},
I.G.~Roseboom\altaffilmark{19},
M.~Rowan-Robinson\altaffilmark{16},
B.~Schulz\altaffilmark{3,28},
Douglas~Scott\altaffilmark{14},
K.~Scott\altaffilmark{5},
P.~Serra\altaffilmark{2},
N.~Seymour\altaffilmark{29},
D.L.~Shupe\altaffilmark{3,28},
A.J.~Smith\altaffilmark{19},
M.~Symeonidis\altaffilmark{29},
K.E.~Tugwell\altaffilmark{29},
M.~Vaccari\altaffilmark{20},
E.~Valiante\altaffilmark{14},
I.~Valtchanov\altaffilmark{35},
A.~Verma\altaffilmark{15},
M.P.~Viero\altaffilmark{3},
L.~Vigroux\altaffilmark{22},
L.~Wang\altaffilmark{19},
D.~Wiebe\altaffilmark{14},
G.~Wright\altaffilmark{23},
C.K.~Xu\altaffilmark{3,28},
G.~Zeimann\altaffilmark{36},
M.~Zemcov\altaffilmark{3,10},
and J.~Zmuidzinas\altaffilmark{3,10}}
\altaffiltext{1}{Center for Astrophysics and Space Astronomy, 389 UCB, University of Colorado, Boulder, CO 80309}
\altaffiltext{2}{Dept. of Physics \& Astronomy, University of California, Irvine, CA 92697}
\altaffiltext{3}{California Institute of Technology, 1200 E. California Blvd., Pasadena, CA 91125}
\altaffiltext{4}{Institute of Astronomy, University of Cambridge, Madingley Road, Cambridge CB3 0HA, UK}
\altaffiltext{5}{Department of Physics and Astronomy, University of Pennsylvania, Philadelphia, PA 19104}
\altaffiltext{6}{Cardiff School of Physics and Astronomy, Cardiff University, Queens Buildings, The Parade, Cardiff CF24 3AA, UK}
\altaffiltext{7}{Department of Physics and Astronomy, Rutgers, The State University of New Jersey, 136 Frelinghuysen Rd, Piscataway, NJ 08854}
\altaffiltext{8}{Institut d'Astrophysique Spatiale (IAS), b\^atiment 121, Universit\'e Paris-Sud 11 and CNRS (UMR 8617), 91405 Orsay, France}
\altaffiltext{9}{Harvard-Smithsonian Center for Astrophysics, 60 Garden Street, Cambridge, MA 02138}
\altaffiltext{10}{Jet Propulsion Laboratory, 4800 Oak Grove Drive, Pasadena, CA 91109}
\altaffiltext{11}{Space Science Building, Cornell University, Ithaca, NY, 14853-6801}
\altaffiltext{12}{Laboratoire d'Astrophysique de Marseille, OAMP, Universit\'e Aix-marseille, CNRS, 38 rue Fr\'ed\'eric Joliot-Curie, 13388 Marseille cedex 13, France}
\altaffiltext{13}{Laboratoire AIM-Paris-Saclay, CEA/DSM/Irfu - CNRS - Universit\'e Paris Diderot, CE-Saclay, pt courrier 131, F-91191 Gif-sur-Yvette, France}
\altaffiltext{14}{Department of Physics \& Astronomy, University of British Columbia, 6224 Agricultural Road, Vancouver, BC V6T~1Z1, Canada}
\altaffiltext{15}{Department of Astrophysics, Denys Wilkinson Building, University of Oxford, Keble Road, Oxford OX1 3RH, UK}
\altaffiltext{16}{Astrophysics Group, Imperial College London, Blackett Laboratory, Prince Consort Road, London SW7 2AZ, UK}
\altaffiltext{17}{Institut de RadioAstronomie Millim\'etrique, 300 Rue de la Piscine, Domaine Universitaire, 38406 Saint Martin d'H\`eres, France}
\altaffiltext{18}{Dept. of Astrophysical and Planetary Sciences, CASA 389-UCB, University of Colorado, Boulder, CO 80309}
\altaffiltext{19}{Astronomy Centre, Dept. of Physics \& Astronomy, University of Sussex, Brighton BN1 9QH, UK}
\altaffiltext{20}{Dipartimento di Astronomia, Universit\`{a} di Padova, vicolo Osservatorio, 3, 35122 Padova, Italy}
\altaffiltext{21}{NRAO, PO Box 2, Green Bank, WV 24944}
\altaffiltext{22}{Institut d'Astrophysique de Paris, UMR 7095, CNRS, UPMC Univ. Paris 06, 98bis boulevard Arago, F-75014 Paris, France}
\altaffiltext{23}{UK Astronomy Technology Centre, Royal Observatory, Blackford Hill, Edinburgh EH9 3HJ, UK}
\altaffiltext{24}{Institute for Astronomy, University of Edinburgh, Royal Observatory, Blackford Hill, Edinburgh EH9 3HJ, UK}
\altaffiltext{25}{Centre for Astrophysics Research, University of Hertfordshire, College Lane, Hatfield, Hertfordshire AL10 9AB, UK}
\altaffiltext{26}{ Institute of Cosmology and Gravitation, University of Portsmouth, Dennis Sciama Building, Burnaby Road, Portsmouth PO1 3FX, UK}
\altaffiltext{27}{Institute for Space and Astronautical Science, Japan Aerospace and Exploration Agency, Sagamihara, Kana- gawa 229-8510, Japan}
\altaffiltext{28}{Infrared Processing and Analysis Center, MS 100-22, California Institute of Technology, JPL, Pasadena, CA 91125}
\altaffiltext{29}{Mullard Space Science Laboratory, University College London, Holmbury St. Mary, Dorking, Surrey RH5 6NT, UK}
\altaffiltext{30}{Space Science \& Technology Department, Rutherford Appleton Laboratory, Chilton, Didcot, Oxfordshire OX11 0QX, UK}
\altaffiltext{31}{Institute for Space Imaging Science, University of Lethbridge, Lethbridge, Alberta, T1K 3M4, Canada}
\altaffiltext{32}{Instituto de Astrof{\'\i}sica de Canarias (IAC), E-38200 La Laguna, Tenerife, Spain}
\altaffiltext{33}{Departamento de Astrof{\'\i}sica, Universidad de La Laguna (ULL), E-38205 La Laguna, Tenerife, Spain}
\altaffiltext{34}{Hubble Fellow}
\altaffiltext{35}{Herschel Science Centre, European Space Astronomy Centre, Villanueva de la Ca\~nada, 28691 Madrid, Spain}
\altaffiltext{36}{University of California, 1 Shields Ave, Davis, CA 95616}

\shortauthors{Conley \etal}
\email{ alexander.conley@colorado.edu }

\begin{abstract}
We report the discovery of a bright ($f\left( 250\mum \right) > 400$
mJy), multiply-lensed submillimeter galaxy \obj\ in {\it
  Herschel}/SPIRE Science Demonstration Phase data from the HerMES
project \citep{Oliver:2010}.  Interferometric 880\mum\ Submillimeter
Array observations resolve at least four images with a large separation of
$\sim 9\arcsec$.  A high-resolution adaptive optics $K_p$ image with
Keck/NIRC2 clearly shows strong lensing arcs.  Follow-up spectroscopy
gives a redshift of $z=2.9575$, and the lensing model gives a total
magnification of $\mu \sim 11 \pm 1$.  The large image separation
allows us to study the multi-wavelength spectral energy distribution
(SED) of the lensed source unobscured by the central lensing mass.
The far-IR/millimeter-wave SED is well described by a modified blackbody fit
with an unusually warm dust temperature, $88 \pm 3$ K.  We derive a
lensing-corrected total IR luminosity of $\left(1.43 \pm 0.09\right)
\times 10^{13}\, \mathrm{L}_{\odot}$, implying a star formation rate of
$\sim 2500\, \mathrm{M}_{\odot}\, \mathrm{yr}^{-1}$.  However, models
primarily developed from brighter galaxies selected at longer
wavelengths are a poor fit to the full optical-to-millimeter SED. A number of
other strongly lensed systems have already been discovered in early
{\it Herschel} data, and many more are expected as additional data are
collected.
\end{abstract}

\section{INTRODUCTION}
\nobreak
\label{sed:introduction}
The discovery of a population of high-redshift, prodigiously
star-forming galaxies at sub-millimeter wavelengths has revolutionized our
understanding of cosmological star formation
\citep[e.g.,][]{2002PhR...369..111B, 2005ApJ...622..772C}.  These
submillimeter galaxies (SMGs) are frequently faint at optical
wavelengths due to significant extinction, but some have far-infrared
luminosities in excess of $10^{13}\, \mathrm{L}_{\odot}$, and are
forming stars at $> 1000\,\mathrm{M}_{\odot}\, \mathrm{yr}^{-1}$.
They are believed to be the progenitors of nearby massive elliptical
galaxies \citep{2008MNRAS.391..420S}, yet many of their properties
remain mysterious.

Dusty, star forming galaxies are responsible for most of the cosmic
infrared background \citep[CIB, e.g.,][]{2009ApJ...707.1729M,
  2010MNRAS.409..109G}, which contains as much energy as all of the
optical light ever emitted by galaxies \citep{1996A&A...308L...5P}.
Modelers have had some success in fitting the spectral energy
distributions (SEDs) of SMGs and using this to infer their properties
\citep[e.g.,][]{2008MNRAS.386..697R}, but it is difficult to study the
sources that produce the CIB in detail because they are individually
faint.  These efforts are biased towards extremely luminous, red
galaxies by selection effects, so it is interesting to test how well
such models describe the less luminous {\it Herschel} sources selected
at shorter wavelengths.

Confusion noise generally sets the flux limit at which individual {\it
  Herschel} sources can be studied.  Gravitational lensing allows this
limit to be circumvented.  Due to the rapidly rising source counts at
faint flux densities \citep{2010MNRAS.409..109G} and the negative
$K$-correction at sub-millimeter wavelengths, strong lensing is
expected to be relatively common for SMGs \citep{1996MNRAS.283.1340B}.
Indeed, follow up of early SPIRE data has shown that a large fraction
of the brightest sources are lensed by other galaxies
\citep{2010Sci...330..800N}.  \citet{2010ApJ...719..763V} discovered a
population of bright galaxies at 1.4 and 2mm, and suggested that these
are lensed.  Lensing allows us to study the properties of
intrinsically fainter SMGs at a level of detail that is currently
difficult otherwise \citep{2010Natur.464..733S,2010A&A...518L..35I}.
Galaxy-galaxy lensing is expected to dominate, with generally small
image separations, so emission and absorption associated with the
foreground lens may obscure the SMG in the optical and near-IR; this
is the case for all of the sources in \citet{2010Sci...330..800N}.

Here we report the discovery of an SMG system (\obj , hereafter
\shortobj ) multiply lensed by a group of galaxies at $z=2.9575 \pm
0.0001$ in Science Demonstration Phase {\it Herschel}/SPIRE
observations of the Lockman-SWIRE field as part of the {\it Herschel}
Multi-tiered Extragalactic Survey (HerMES; S.~Oliver et al.\ 2011, in
preparation), with coordinates $\alpha =
10\mathrm{h}57\mathrm{m}51\mathrm{s}$
$\delta=57^{\circ}30\arcmin27\arcsec$ (J2000).  A number of additional
lensed systems are already known in HerMES data.  The large separation
between the images allows us to measure the SED of this object across
a long wavelength baseline. In this letter, we model the optical-to-millimeter
SED of this object. The lensing model for this system, based on
high-resolution optical and near-IR observations, is presented in
\citet[][hereafter G11]{Gavazzi:11}. We have also obtained
high-resolution CO line maps \citep[][hereafter R11]{Riechers:11},
and used the CO line strength distribution to model the molecular gas
\citep[][hereafter S11]{Scott:11}.

\section{OBSERVATIONS}
\label{sec:observations}
\nobreak
 \shortobj\ was discovered using observations with the Spectral and
Photometric Imaging Receiver \citep[SPIRE,][]{2010A&A...518L...3G}
on-board {\it Herschel} \citep{2010A&A...518L...1P}.  It was selected
for further follow-up with Z-Spec, a millimeter-band grating spectrograph at
the Caltech Submillimeter Observatory \citep{2006SPIE.6275E..32E}, based on
its brightness and blue color ($f\left(500\mum\right) <
f\left(300\mum\right)$); the latter was intended to avoid $z > 4$
sources where a redshift would be difficult to obtain.  Z-Spec gives a
secure redshift of $z = 2.958 \pm 0.007$ (S11).  R11 use additional CO
lines measured with the Plateau de Bure Interferometer (PdBI), the
Combined Array for Research in Millimeter-wave Astronomy (CARMA), and
Zpectrometer on the Green Bank Telescope to refine the redshift,
yielding $z=2.9575 \pm 0.0001$.

\shortobj\ is unresolved in the diffraction-limited SPIRE observations
($\mathrm{FWHM}_{250\mum} = 18.6\arcsec$).  The extreme brightness
($f\left( 250\mum \right) \simeq 400\, \mathrm{mJy}$) and the
morphology in the optical and near-IR suggested a lensed source.  A
Subaru $i$ image from the SERVS
survey\footnote{\url{http://www.cv.nrao.edu/~mlacy/servs.html}} shows
clear evidence of lensing.  The source is visible in previously
obtained $gr$ WHT observations and in data from the {\it
  Spitzer}/SWIRE survey (Surace et al.\ 2011, in preparation).  We
obtained observations with the Submillimeter Array (SMA) in compact
configuration at 880\mum\ (beam FWHM 2.3\arcsec ), resolving the
source into at least four components matching the position of several
optical sources and surrounding a foreground elliptical galaxy.  We
further obtained a $K_p$ adaptive optics (AO) observation using NIRC2
on the KeckII telescope and the laser guide-star system
\citep{2006PASP..118..297W}.  PdBI CO maps, presented in R11, detect
at least the two brightest sources and show that they are at the same
redshift.  The photometric redshift of the central elliptical is $0.60
\pm 0.04$ \citep{Oyaizu:2008}.  An optical spectrum obtained with the
double spectrograph on the Hale telescope has absorption features
consistent with the photo-$z$.

\begin{figure}
\plotone{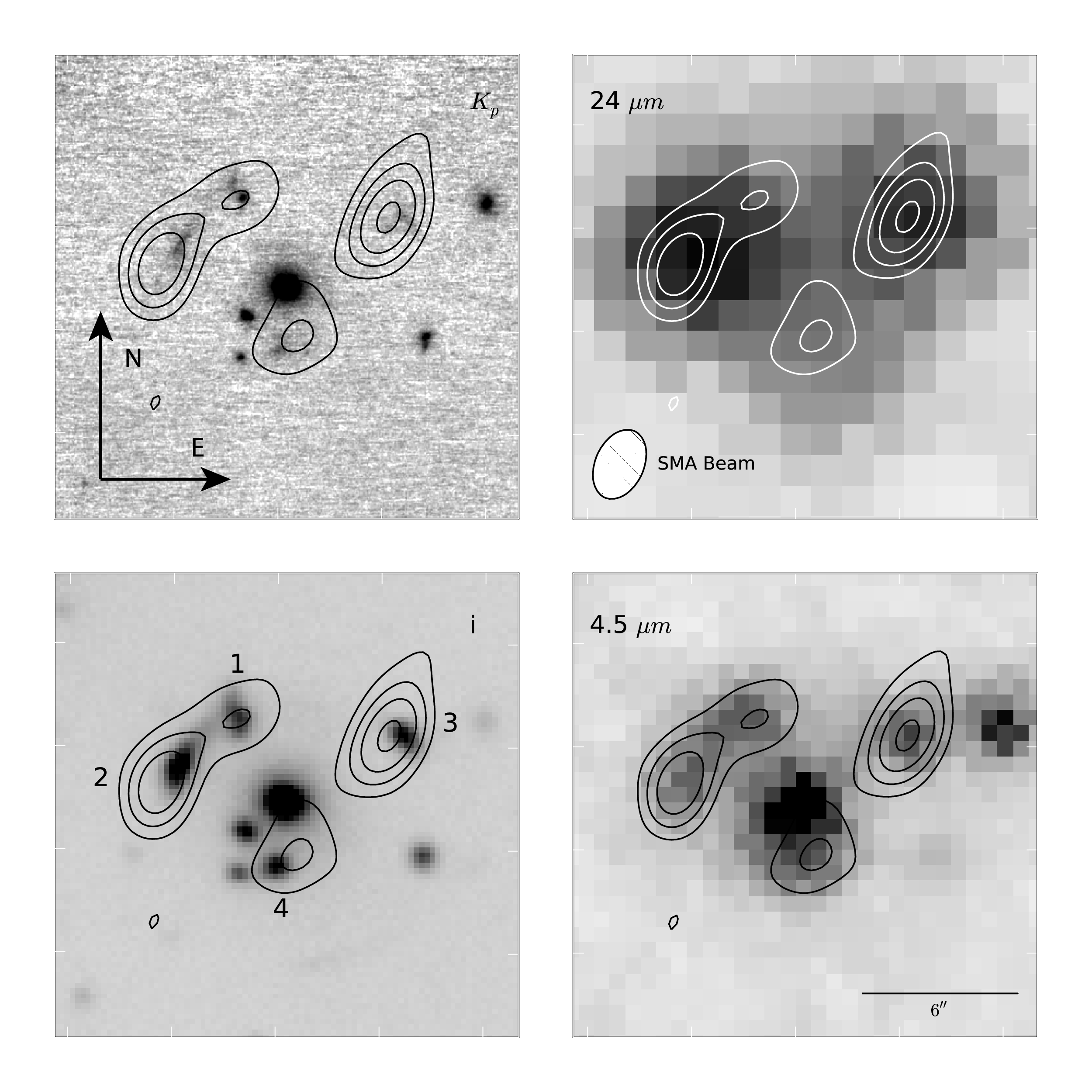}
\caption{A composite of selected multi-wavelength observations of
  \shortobj .  Each image is $18\arcsec \times 18\arcsec$.
  Clockwise from the upper left: Keck $K_p$ AO image; {\it
    Spitzer}/MIPS 24\mum ; {\it
    Spitzer}/IRAC 4.5\mum ; Subaru SuprimeCam
  $i$.  In all panels the contours show the SMA 880\mum\ 
  observations.  The numbering scheme used to identify sources
  in this letter is shown in the bottom left panel.  The highest
  resolution {\it Herschel}/SPIRE passband has a resolution of
  18.6\arcsec , larger than the entire field shown
  here. \label{fig:multicolor} }
\end{figure}

The numbering scheme used to identify sources in this letter is shown
in Figure~\ref{fig:multicolor}, along with as a montage of
observations at multiple wavelengths.  The high-resolution $K_p$ image
was used to construct a lensing model, which was compared with the $i$
and PdBI observations to constrain differential magnification; see G11 for
details.  This model has five lensing masses at the locations of
foreground galaxies, and gives a total magnification factor of $\mu =
10.86 \pm 0.68$ for all five detected images based on a
cored-isothermal model.  The velocity dispersion strongly suggests
that the lenses reside in a massive group of galaxies. 

The photometry is summarized in Table~\ref{tbl:fluxes}.  At optical
wavelengths the individual images are blended and partially resolved,
so neither point-spread-function (PSF) nor aperture photometry is
entirely satisfactory.  We use aperture photometry with a relatively
small aperture radius to minimize blending, and compute the aperture
corrections for the partially resolved sources by convolving the lens
model to the matching resolution in each band; the adjustment to the
aperture correction compared with isolated point sources is only a few
percent.  We exclude photometry of image 1 because it is contaminated
by a foreground object, and again use the lensing model to correct for
the omitted light, which is a $\sim 15\%$ correction.

In the {\it Spitzer}/SWIRE data, the individual images are separable
in the Infrared Array Camera (IRAC) bands (3.6 to 8\mum), and the
SWIRE PSF photometry is adequate for our purposes.  We use only
photometry of sources 2 and 3, again using the lensing model to
correct for the omitted sources 1 and 4.  \shortobj\ is not present in
the SWIRE catalog at 3.6\mum .  The FWHM of the 70 and 160\mum\ MIPS
observations are much larger than the separation between individual
images, so the catalog flux measurement already includes all the
images.  At 24\mum\ the SWIRE aperture is just smaller than the
separation, so we re-measured the photometry using a larger aperture.

The SPIRE fluxes used here are from the HerMES SCATv3.1 catalog
(A.~J. Smith et al.\ 2011, in preparation).  For the SMA data we
extract photometry and positions using the
CASA\footnote{http://casa.nrao.edu} {\tt imfit} task. For Z-Spec, we
bin the spectrum into 5 100\mum\ bins after masking noisy channels and
the detected CO lines.  The 20cm photometry is from the FIRST survey
\citep{1995ApJ...450..559B}.

We place upper limits on the potential foreground contamination by
scaling the observed elliptical galaxy SEDs from
\citet{2007ApJ...655..863D, 2007ApJ...660.1215T} to match the optical
magnitudes of the central lensing elliptical, taking the highest
resulting fluxes in each band as our contamination limit.  This is not
relevant at shorter wavelengths where the sources are clearly
resolved.  The potential contamination peaks at 6 mJy at 160\mum , and
3 mJy at 70\mum .  Given the SED, only the potential
70\mum\ contamination is significant, but we adopt the contamination
limits as an additional correlated uncertainty at all wavelengths.  We
take calibration errors, which also affect the photometry in a
correlated fashion, from the instrument documentation.

\begin{deluxetable}{lll}
\tablewidth{0pt}
\tablecaption{Photometry}
\tablehead{
 \colhead{ Wavelength [\mum ]} &
 \colhead{ Flux Density } & \colhead{Telescope/Detector} 
}
\startdata
0.48 ($g$) & $26.0 \pm 0.3$ $\mu$Jy  & INT/WFC\\
0.63 ($r$) & $46.5 \pm 0.6$ $\mu$Jy  & INT/WFC\\
0.76 ($i$) & $47.9 \pm 0.3$ $\mu$Jy  & Subaru/SuprimeCam\\
2.2  ($K_p$) & $63.1 \pm 2.0$ $\mu$Jy  & KeckII/NIRC2\\
4.5        & $376 \pm 6$ $\mu$Jy     & {\it Spitzer}/IRAC\\
5.8        & $442 \pm 11$ $\mu$Jy     & {\it Spitzer}/IRAC\\
8.0        & $558 \pm 16$ $\mu$Jy    & {\it Spitzer}/IRAC\\
24         & $5.5 \pm 0.4$ mJy       & {\it Spitzer}/MIPS\\
72         & $22.2 \pm 3$ mJy          & {\it Spitzer}/MIPS\\
160        & $310 \pm 8$ mJy         & {\it Spitzer}/MIPS\\
250        & $425 \pm 10$ mJy        & {\it Herschel}/SPIRE\\
350        & $340 \pm 10$ mJy        & {\it Herschel}/SPIRE\\
510        & $233 \pm 11$ mJy        & {\it Herschel}/SPIRE\\
880        & $52.8 \pm 0.5$ mJy      & SMA \\
1000-1100  & $27.5 \pm 0.6$ mJy      & CSO/Z-Spec\\
1100-1200  & $20.4 \pm 0.5$ mJy      & CSO/Z-Spec\\
1200-1300  & $16.2 \pm 0.5$ mJy      & CSO/Z-Spec\\
1300-1400  & $12.0 \pm 0.5$ mJy      & CSO/Z-Spec\\
1400-1500  & $9.9 \pm 0.6$ mJy       & CSO/Z-Spec\\
3400       & $0.61 \pm 0.19$ mJy     & CARMA\\
214000     & $1.8 \pm 0.7$ mJy       & VLA\\
\enddata
\tablecomments{Combined flux densities for all the detected images as
  detailed in the text.  For brevity, only the summed flux density from
  all images is provided.  All values are calibrated relative to a
  $F_{\nu} = \mathrm{const}$ SED.  Note that calibration errors, which
  are dominant at most wavelengths and are strongly correlated between
  points, are not included, and neither are contamination/confusion
  errors. These values have not been corrected for magnification.}
\label{tbl:fluxes}
\end{deluxetable}

\section{SED fitting and Source Properties}
\label{sec:templates}
\nobreak 
We first analyze the SED by fitting simple modified blackbody models
to the long-wavelength data.  It is possible that the far-IR emission
has a different spatial distribution than the shorter wavelength data
used to derive the lensing model, but the current SMA observations are
not high enough resolution to explore this issue.  Therefore, we
assume that the near- and far-IR emission are co-located.  G11 find
evidence for a small shift ($\sim 0.4\arcsec$) between the PdBI CO
emission and the optical emission, too small to significantly affect
our results.

The standard form for a modified blackbody SED is a
frequency-dependent optical depth factor: $f_{\nu} \propto \left( 1 -
\exp \left[ -\tau \left( \nu \right) \right] \right) B_{\nu} \left( T
\right)$, where $B_{\nu}$ is the Planck function.  The optical depth
is assumed to be a power law in frequency, $\tau = \left(\nu / \nu_0
\right)^{\beta}$ following \citet{2006ApJ...636.1114D}.  $\lambda_0 =
c / \nu_0$ is the wavelength where the optical depth is unity.  In the
optically thin case, $\nu \ll \nu_0$, this reduces to $f_{\nu} \propto
\nu^{\beta} B_{\nu} \left( T \right)$.  The latter is often used in
the literature to estimate temperatures, but here our data allow us
to drop the assumption of optical thinness.  In both cases we join the
modified blackbody to a simple power law on the blue side of the peak
\citep{2003MNRAS.338..733B}, which only affects the
70\mum\ observation.

\begin{figure}
\plotone{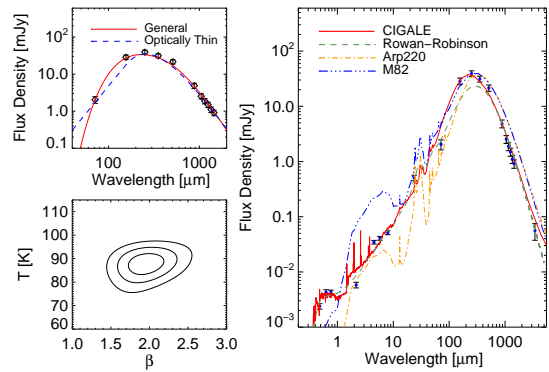}
\caption{SED fits to \shortobj\ after correction for flux
  magnification.  The left-hand panels show the modified blackbody
  fits to the long-wavelength observations (top panel) and the
  constraints on the temperature and $\beta$ (the contours correspond
  to 68/95/99\% enclosed probability) for the general form of the
  model.  Note that the data points are strongly correlated, and the
  optically thick fit is an excellent match to the data.  The right
  hand panel compares the optical to millimeter SED compared with the
  best fitting {\tt CIGALE} model, the best combination of templates
  from \citet{2010arXiv1006.0120R}, and the IR SEDs of Arp220 and M82
  (prototypical nearby IR-luminous galaxies) scaled to have the same
  maximum flux density.  All wavelengths are observer frame.
\label{fig:seds}}
\end{figure}

We fit both models from $70\mum$ to 1.5mm to derive the temperature
and total IR luminosity, including the error on the magnification.
The fit is shown in the left hand panels of Figure~\ref{fig:seds}, and
the parameters are given in Table~\ref{tbl:greybodyparams}.  All of
the parameters are well-constrained by our data.  We find $\lambda_0 
\simeq 200\mum$, a reasonable match to the theoretically
expected value $\lambda_0 \simeq 100\mum$ \citep{2006ApJ...636.1114D}
and similar to that derived for Arp 220 \citep{2003MNRAS.338..733B}.
The emission is optically thick bluer than observer-frame $\sim
800\mum$.  A two-temperature model decreases the $\chi^2$ by $< 0.002$
for the optically thick model (since there is virtually no
contribution from the second component for the best fit), and for the
optically thin model by about $5$, so the latter remains a very
poor fit.

The temperatures and $\beta$ values for the two different models
disagree strongly; this is also the case for the fits to Arp 220 in
\citet{2003MNRAS.338..733B}.  The general model (i.e., $1-\exp\left[ -
  \tau\right]$) fits our data quite well, but the optically thin model
does not, with a reduced $\chi^2_{\nu} > 3$.  The derived temperature
for the former is fairly high ($\sim 90$ K), suggesting a
dust-enshrouded active galactic nucleus (AGN) contribution.  This
result is robust against removing the data from any single instrument
or any individual data point.  Obtaining an acceptable fit for a more
typical dust temperature, such as 60K, requires increasing all the
errors by a factor of $>2.2$, including calibration errors.  It is
possible that the lensing is selectively magnifying a warm component,
and that this high temperature is not representative of the SMG as a
whole. The poor quality of the optically thin fit is similarly robust,
unless the SPIRE observations are removed, in which case it becomes
acceptable ($\chi^2_{\nu} = 1.1$).  It is therefore possible that such
warm SMGs have been missed in previous surveys that did not have
observations near the peak of the SED, as it would then be difficult
to distinguish between the optically-thin and thick cases.  Since the
general model fits the data much better, and makes fewer assumptions,
henceforth we only discuss the results of this fit.

The above findings are independent of the lensing model, unless the
location of the emission (and hence the magnification) changes
strongly from 250\mum\ to 1.5mm, which is unlikely.  Turning to
quantities which must be corrected for the lensing magnification, we
find a far-IR luminosity of $\mathrm{L}_{\mathrm{IR}} = 1.43 \times
10^{13}\, \mathrm{L}_{\odot}$, where $\mathrm{L}_{\mathrm{IR}}$ is
defined as the luminosity from $8$ to $1000\mum$ in the rest frame.
This implies a star formation rate of $\sim 2500\,
\mathrm{M}_{\odot}\, \mathrm{yr}^{-1}$ from the relation of
\citet{1998ApJ...498..541K}, ignoring any AGN contribution.  We also
measured $\mathrm{L}_{\mathrm{IR}}$ by spline-interpolating the
observations, which gives a similar value ($1.49 \times 10^{13}\,
\mathrm{L}_{\odot}$).  Assuming a mass-absorption coefficient of
$\kappa_{\nu} = 2.64\, \mathrm{m}^2 \mathrm{kg}^{-1}$ at
125\mum\ following \citet{2003MNRAS.341..589D}, from the temperature
and luminosity we infer a dust mass of $M_d \simeq 1 \times 10^8\,
\mathrm{M}_{\odot}$; $\kappa_{\nu}$, and hence $M_d$, is uncertain by
at least a factor of three.  Further assuming a molecular gas to dust
ratio of 60 for SMGs \citep{2008MNRAS.389...45C}, we estimate a gas
depletion time of $\sim 2.4 \times 10^6\, \mathrm{yr}$, considerably
shorter than the value of $> 4 \times 10^7\, \mathrm{yr}$ for
`typical' SMGs derived by e.g., \citet{2010ApJ...714L.118D}; see R11
for actual gas mass estimates.

\begin{deluxetable}{lll}
\tablecaption{Modified Blackbody Fits}
\tablehead{
 \colhead{Model} & \colhead{General: $1-e^{-\tau}$} & \colhead{Optically Thin:
  $\nu^{\beta}$ }
}
\startdata
T & $88.0 \pm 2.9$ K & $48.5 \pm 2.6$ K \\
$\beta$ & $1.95 \pm 0.14$ & $1.61 \pm 0.15$ \\
$\lambda_0$ & $197 \pm 19$ \mum & NA \\
\hline
$\mathrm{L}_{\mathrm{IR}}$ & 
  $\left(1.43 \pm 0.09\right) \times 10^{13}\, \mathrm{L}_{\odot}$ & 
 $\left(1.13 \pm 0.09\right) \times 10^{13}\, \mathrm{L}_{\odot}$ \\
SFR   & $2460 \pm 160\, \mathrm{M}_{\odot} \mathrm{yr}^{-1}$ &
 $1950 \pm 160\, \mathrm{M}_{\odot} \mathrm{yr}^{-1}$  \\
$\chi^2$ & 6.77 for 6 dof & 22.1 for 7 dof \\ 
\enddata
\tablecomments{Fit values for the two modified blackbody models,
 applied to the magnification corrected
 $70 - 1500\mum$ photometry. The second model assumes optically
 thin emission, and is only presented for comparison
 with literature values. The derived parameters include the uncertainty
 in the magnification.  For $L_{\mathrm{IR}}$, we assume 
 $h=0.7, \Omega_m = 0.27,$ and $\Omega_{\Lambda} = 0.73$.
\label{tbl:greybodyparams}}
\end{deluxetable}

This source is weakly detected at 1.4 GHz in the FIRST survey, so we
also compute $q_{\mathrm{IR}}$, the logarithmic ratio of
$L_{\mathrm{IR}}$ and the rest-frame 1.4 GHz flux density.
\shortobj\ has an moderately low ratio, $q_{\mathrm{IR}} = 1.5 \pm
0.2$, compared with the mean value and scatter of $q_{\mathrm{IR}} =
2.4 \pm 0.12$ for HerMES sources with firm radio cross-identifications
\citep{2010A&A...518L..31I}, although not the lowest found.  The high
1.4 GHz flux density suggests some AGN contribution.  However, the
5\arcsec\ resolution of the FIRST survey is not good enough to rule
out foreground contamination.  While this is likely small,
$q_{\mathrm{IR}}$ should be regarded as a lower limit.

In addition to the above simple models, we have also investigated a
variety of template fits across all wavelengths using several
packages.  A similar study was carried out for isolated, but somewhat
intrinsically brighter, {\it Herschel} sources by
\citet{2010MNRAS.409...66B}, who found that model templates were
generally a good match to the data.  This is not the case here;
available templates generally do not fit both the near-IR and far-IR
through millimeter-wave data simultaneously.  The templates of
\citet{2008MNRAS.386..697R,2010arXiv1006.0120R} underestimate the
far-IR flux by $\sim 30\%$.  This is also the case for the models of
\citet{2007A&A...461..445S}.  In both cases this is because the SED of
\shortobj\ peaks blueward of the templates.  Similarly, the models of
\citet{2008MNRAS.388.1595D} are unable to reproduce the full SED
(E. da~Cunha 2010, private communication).  The potential foreground
contamination discussed in Section~2 is too small to explain these
issues.

Next we turn to the {\tt CIGALE} package \citep{2009A&A...507.1793N},
which combines optical/near-IR templates with a longer-wavelength dust
model.  Using a two-stellar population model, this fits the SED
considerably better, but has some issues in the near-IR,
overpredicting $K_p$ (rest frame $g$) flux density by a factor of
two and missing the slope of the IRAC observations.  We are unable to
explain this discrepancy.  Ignoring these issues, {\tt CIGALE} finds a
total stellar mass of $\log_{10} \mathrm{M}_{*} = 10.8^{+0.2}_{-0.3}$,
and a star formation rate of $\log_{10} \mathrm{SFR} =
3.3^{+0.4}_{-0.5}$, in $\mathrm{M}_{\odot}$ and $\mathrm{M}_{\odot}\,
\mathrm{yr}^{-1}$, respectively.  $70 \pm 30\%$ of the stars are in a
young, strongly extinguished stellar component ($A_V = 6 \pm 2$),
whose age is not well constrained.  A sample of SED fits is shown in
Figure~\ref{fig:seds}.

\section{Conclusions}
\nobreak 
The unusually large image separation of \shortobj\ compared with
most lensed sub-millimeter sources provides an opportunity to study a
sub-millimeter galaxy at a level of detail typically only possible
for more luminous sources.  Due to confusion noise, this will
generally not be feasible for un-lensed sources until the completion
of ALMA.  Detailed models of the gas and dust content based on CO
emission are presented in S11 and R11.

Modified blackbody fits to the long-wavelength data (70\mum\ to 1.5mm)
imply a warm dust temperature of $90\, \mathrm{K}$, and a star
formation rate of $\sim 2500\, \mathrm{M}_{\odot}\, \mathrm{yr}^{-1}$.
Compared with other SMGs, we find a short gas depletion timescale of
only a few million years, assuming negligible AGN contribution to
$L_{\mathrm{IR}}$.  This is one of the few SMGs that have been studied
across such a wide wavelength range, due to the large image
separation, so it is interesting that SED fits from the optical to the
millimeter are generally a fairly poor fit to the data, typically matching the
short wavelength data well but underpredicting the far-IR peak.  We
obtain somewhat better results with {\tt CIGALE}, but it significantly
overpredicts the 2\mum\ flux and is not a great match to the IRAC
observations.  It is unclear if this galaxy is simply unusual, or if
the templates --- which were largely derived from even brighter
galaxies selected at longer wavelengths --- are not a good
representation of galaxies selected at wavelengths near the peak of
the far-infrared background.  The latter would have significant
implications for the inferred history of high-$z$ star formation.
Such warm systems may have been missed by previous surveys lacking
data near the peak of the SED because of the common assumption of
optical thinness.  Models suggest there should be a large number of
strongly lensed SMGs in {\it Herschel} data, which is consistent with
early observations \citep{2010Sci...330..800N}, so we expect to
address this question soon.  The variations in lensing magnification
make this a promising tool to study SMGs across a wide range of
intrinsic luminosities, although relatively few sources will have such
a large image separation.  Additional multi-wavelength observations,
particularly at high resolution, would improve the SED constraints
significantly.

\acknowledgements SPIRE has been developed by a consortium of
institutes led by Cardiff Univ.\ (UK) and including Univ.\ Lethbridge
(Canada); NAOC (China); CEA, LAM (France); IFSI, Univ.\ Padua (Italy);
IAC (Spain); Stockholm Observatory (Sweden); Imperial College London,
RAL, UCL-MSSL, UKATC, Univ.\ Sussex (UK); Caltech, JPL, NHSC,
Univ.\ Colorado (USA).  This development has been supported by
national funding agencies: CSA (Canada); NAOC (China); CEA, CNES, CNRS
(France); ASI (Italy); MCINN (Spain); SNSB (Sweden); STFC (UK); and
NASA (USA).  The Submillimeter Array is a joint project between the
Smithsonian Astrophysical Observatory and the Academia Sinica
Institute of Astronomy and Astrophysics.  The IRAM Plateau de
Bure Interferometer is supported by INSU/CNRS (France), MPG (Germany)
and IGN (Spain).  The National Radio Astronomy Observatory is a
facility of the National Science Foundation operated by Associated
Universities, Inc.  Support for CARMA construction was derived from
the Gordon and Betty Moore Foundation, the Kenneth T. and Eileen
L. Norris Foundation, the James S. McDonnell Foundation, the
Associates of the California Institute of Technology, the University
of Chicago, the states of California, Illinois, and Maryland, and the
National Science Foundation. Ongoing CARMA development and operations
are supported by by NSF grant ATI-0838178 to CARMA, and by the CARMA
partner universities.  The authors would like to thank Elisabete
da~Cunha for running her models for us.  The {\it Herschel} data
presented in this letter will be released through the {\em Herschel}
Database in Marseille, HeDaM\footnote{\url{hedam.oamp.fr/HerMES}}.

{\it Facilities:} \facility{{\it Herschel} (SPIRE)}, \facility{CSO
  (Z-Spec)}, \facility{Subaru (SuprimeCam)}, \facility{SMA},
\facility{Hale (SWIFT)}, \facility{Keck:II (NIRC2)}, \facility{IRAM:
  Interferometer}, \facility{ING: Newton (WFC)}, \facility{{\it
    Spitzer} (IRAC)}, \facility{{\it Spitzer} (MIPS)},
\facility{CARMA}, \facility{VLA}, \facility{GBT (Zpectrometer)}


\end{document}